\documentclass[bibyear]{aa}
\usepackage{enumitem}\setlist[itemize,1]{label=$\diamond$}
\usepackage{placeins}
\usepackage{graphicx}
\usepackage{multirow}
\usepackage{txfonts}
\usepackage{siunitx}
\usepackage{subcaption}
\usepackage{natbib,twoopt}
\usepackage[breaklinks=true]{hyperref}
\bibpunct{(}{)}{;}{a}{}{,}
\makeatletter
  \newcommandtwoopt{\citeads}[3][][]{\href{http://adsabs.harvard.edu/abs/#3}%
    {\def\hyper@linkstart##1##2{}%
     \let\hyper@linkend\@empty\citealp[#1][#2]{#3}}}
  \newcommandtwoopt{\citepads}[3][][]{\href{http://adsabs.harvard.edu/abs/#3}%
    {\def\hyper@linkstart##1##2{}%
     \let\hyper@linkend\@empty\citep[#1][#2]{#3}}}
  \newcommandtwoopt{\citetads}[3][][]{\href{http://adsabs.harvard.edu/abs/#3}%
    {\def\hyper@linkstart##1##2{}%
     \let\hyper@linkend\@empty\citet[#1][#2]{#3}}}
  \newcommandtwoopt{\citeyearads}[3][][]%
    {\href{http://adsabs.harvard.edu/abs/#3}
    {\def\hyper@linkstart##1##2{}%
     \let\hyper@linkend\@empty\citeyear[#1][#2]{#3}}}
\makeatother

\begin{document} 

   \title{$R_V$ from multi-waveband galaxy polarimetry in supernovae vicinity}

   \author{J. Rino-Silvestre\inst{1}
           \and
           S. Gonz\'alez-Gait\'an\inst{2}
           \and
           A. Mour\~ao\inst{1}
           \and
           J. Duarte\inst{1}
           \and
           B. Pereira\inst{1}
           }
           
   \institute{
        CENTRA, Instituto Superior Técnico, Universidade de Lisboa,
              Av. Rovisco Pais, 1049-001, Lisboa, Portugal\\ \email{joao.silvestre@tecnico.ulisboa.pt}\\
              \email{amourao@tecnico.ulisboa.pt}\\
              \email{joao.r.d.duarte@tecnico.ulisboa.pt}\\
              \email{beatriz.m.pereira@tecnico.ulisboa.pt}\\
        \and
       Instituto de Astrofísica e Ciências do Espaço, Faculdade de Ciências, Universidade de Lisboa, Ed. C8, Campo Grande, 1749-016 Lisbon, Portugal\\
       \email{gongsale@gmail.com}
       }
    
   \date{}

  \abstract
   {Peculiar dust extinction laws have been reported for some type Ia supernovae (SNe) with the parameter $R_V$ much lower than the average value for the Milky Way of 3.1. These findings challenge our understanding of dust properties in galaxies, with unknown implications in cosmology with supernovae.
   }
   {Using optical photopolarimetry of supernova host galaxies, a few years after the explosion, we estimate $R_V$ in the vicinity of each supernova and compare it with the extinction law calculated directly from SN observations.}
   {Multiband photopolarimetric data of nine galaxies, hosts of eleven SNe, acquired with VLT-FORS2 in IPOL mode, are used to map the polarization angle and the polarization degree in each galaxy.  
   Data are processed with a custom-built reduction pipeline that corrects for instrumental, background, and Milky Way interstellar polarization effects. 
   The validity of Serkowski relations is tested at different locations in the galaxy to extract the wavelength of the maximum polarization $\lambda_{max}$ and obtain 2D maps for $R_V$. When the fit to $\lambda_{max}$ at the SN location is poor, or impossible, an approximate Bayesian spatial inference method is employed to obtain an estimate of $\lambda_{max}$ using well-fitted neighboring locations. The estimated local $R_V$ for each SN is compared with published values from the supernova light curves.}
   {We find $R_V$ values from optical photopolarimetry at SNe locations consistent with the average Milky Way value and a median difference of $>3\sigma$ with the low peculiar $R_V$ obtained from the analysis of some reddened SNIa light curves. The $R_V$ estimates obtained with $BVRI$ photopolarimetry for the SNe vicinity are statistically similar to the global $R_v$ we obtain for the host.}
   {The discrepancy between the local $R_V$, inferred from photopolarimetry in the SN vicinity, and $R_V$ obtained from SNe light curves suggests that the extinction laws obtained directly from the SNe may be driven by more local effects, perhaps from the effects of the interaction of the light from the supernova with very nearby material.}


   \maketitle
%

\section{Introduction}\label{sec:intro}
Cosmic dust is ubiquitous in the Universe, particularly present in the interstellar medium \citep[ISM;][]{ismdust} and in the line of sight towards astrophysical objects such as supernovae (SNe) and their remnants \citep{snedust}. Dust grains partially absorb and scatter light in the ultra-violet (UV) / optical wavelengths and re-emit that energy in infrared (IR) and sub-mm wavelengths and are thus responsible for both extinction and reddening of optical light in the line of sight (LOS). Such effects have to be taken into account in distance measurements for cosmology when using "standard candles" such as supernovae \citep[e.g][]{Betoule}.\par

Dust absorption, scattering, and re-emission of light detected from astronomical objects must be accounted for when studying its intrinsic properties, as otherwise incorrect assessments may be made regarding e.g. an object's temperature, luminosity, or distance. Moreover, the nature of the dust can also inform us of physical and chemical processes related to its history, from formation and variation in composition to grain growth and destruction in different astrophysical structures and events, such as accretion disks \citep{add1,add2}, molecular clouds \citep{mcd1} and galaxies \citep{gald1}, as well as its interaction with magnetic fields through dust grain alignment \citep{hoang2016}.\par

Dust can also lead to polarization of the light as it travels through the ISM via two main processes: dust extinction and dust scattering. For the first, the alignment of asymmetric grains in the presence of a magnetic field will be such that their shortest axis is parallel to the field \citep{DGmech, RAT2}. Absorption of the electric-field component parallel to the long axis will be higher than that of the electric-field component parallel to the short axis, leading to a preferential extinction direction. Incident light transmitted in the same direction as the LOS will lead to the observation of polarization due to extinction \citep{hiltner, hall}, the transmitted light being polarized parallel to the magnetic field that aligns the dust grains. The light re-emitted by those grains in the mid/far-infrared will be perpendicularly polarized relative to that field.\par

In galaxy regions where polarization is mostly due to the second process of dust scattering, the scattered flux is expected to follow the Rayleigh law, $1 / \lambda^4$, should the grain size be small in relation to the incident wavelength. However, this also requires the complex refractive index of the dust to be weakly dependent on wavelength \citep{particlesCH5}. Furthermore, the wavelength dependence of the scattered flux will be associated with other factors, such as the chemical composition of the dust and the grain size distribution \citep{VoshFara1993, Fara2005, BCK2022}. Polarization resulting from scattering will be observed in regions around illuminating sources. In the optical wavebands, light that scatters off a medium around a non(minimally)-obscured light source is expected to present an axisymmetric pattern centered on that source, i.e. with the polarization angle $\chi$, perpendicular to the incident light ray (as viewed face-on). This pattern will change depending on the viewing orientation with respect to the scattering plane. Added confusion may be present due to the mixing of both processes, extinction and scattering, when little or no prior information about the dust properties and its geometry is available.\par

In regions where the polarization of light is mostly due to absorption by aligned dust grains, we expect the relationship between the polarization degree and the wavelength of light to follow the empirical Serkowski relation \citep{serkowski}. The maximum wavelength of polarization in this relation is acknowledged to be connected to the size of dust grains in the medium \citep{draine} and has been shown to be linearly correlated with the total-to-selective extinction ratio parameter or "reddening law" $R_V$ \citep{rv, vrba}. 

Dust extinction and possible variations among SN Ia populations are among the largest systemic uncertainties when using SNe Ia for cosmology \citep[e.g.][]{scolnic2014,Brout21,Gonzalez-Gaitan21}. Several studies \citep{fitzpatrick2007, Amanullah2011, Mandel11, Phillips13, carnagie, Amanullah2015, claudia16} have further found that a subset of SNe Ia presents a $R_V$ that departs from the average Milky Way value of 3.1, even going below 2. The most straightforward explanation is simply that peculiar dust properties exist everywhere in the Universe but are rare, and we just happen to observe those in some SNe Ia. In fact, lower $R_V$ values have been observed in some regions of the MW \citep[e.g.][]{Udalski03} and the Magellanic Clouds \citep{Gordon03}.\par
 
Another possibility of the presence of peculiar extinction laws in some SNe~Ia is circumstellar material (CSM) at distances below \SI{1}{pc} ejected by the progenitor prior to explosion \citep{wang2005, goobar2008}. In this scenario, the scattering cross section exceeds the absorption cross section along the visible and NIR wavelengths, and rises at shorter wavelengths, leading to bluer photons reaching the observer and to an enhancement of the selective attenuation, i.e.  a steeper extinction law (lower $R_V$). However,  another prediction of this scenario is a time-dependent strong polarization signal \citep{patat05} that has not been observed so far, although few polarimetric observations during the SN evolution exist to-date \citep[see e.g.][]{kawabata14}.\par

Alternatively, the strong radiative pressure of the explosion could accelerate nearby dust clouds leading to collisions between them \citep{hoang17}. This would lead to i) smaller grains (and lower $R_V$) from collisions, and ii) time-varying $E(B-V)$ and $R_V$ from forward scattering effects, which has indeed been observed \citep[e.g.][]{forster13,bulla18}. Polarization signals would be smaller since the forward scattering angle would also be small \citep{Nagao24}.  

Interestingly, observations of the M82 galaxy in the vicinity of reddened SN 2014J a few years before the explosion, yielded an extinction law indicating average dust properties similar to those of the MW \citep{hutton15} and very different from the peculiar dust population inferred from observations obtained after the explosion from the SN light \citep{Amanullah2014}, again arguing for a nearby material scenario affecting our measurements of $R_V$.\par

The study of optical polarization of extended regions, unlike point sources (SNe), has been scarcely pursued, with only a few examples dealing with galaxies \citep{scarr, scarr2, felton,jones}. In this work we calibrate, reduce, and analyze multi-band polarimetric data of SN host galaxies (HGs) to study the extinction laws at the SN locations, more specifically the local dust within the interstellar medium (ISM) via the interstellar polarization (ISP) it causes. Section \ref{sec:data} describes the data sample used in this work and provides an overview of the reduction methods employed. Section~\ref{sec:models} describes the expected polarization patterns from dust scattering and dust absorption with the help of simple radiative transfer simulations. Section \ref{sec:meth} outlines the analysis and selection of photopolarimetric reduced products. In Section \ref{sec:disc} we present our results for $R_V$ in the vicinity of SNe and compare them against those SNe $R_V$ present in the literature and against the $R_V$ we obtained for the remainder of the HGs. A discussion of their relationship and possible implications follows together with future research routes. 

\section{Data and Reduction Methods}\label{sec:data}
The photopolarimetric data used in this work was obtained in 2017 with the Very Large Telescope's FOcal Reducer and low dispersion Spectrograph (VLT - FORS2)\footnote{\href{https://www.eso.org/sci/facilities/paranal/instruments/fors/overview.html}{ESO VLT/FORS2 Overview}} at the ESO Paranal Observatory through the 099.D-0043 (A/B) program. The data is comprised of multi-waveband ($BVRI$) polarimetric images (and respective calibrations) of a sample of 9 SN host galaxies, identified in Table~\ref{tab:hgid}. These galaxies were chosen because most of their hosted SNe were measured to have particularly low values of $R_V$ \citep{Mandel11, Phillips13, carnagie, Amanullah2015, claudia16}.\par

\begin{table*}[ht]
\centering
\begin{tabular}{|ccc|c|c|cccc|}
\hline
\multicolumn{3}{|c|}{\textbf{Host Galaxy}}                                                                                          & \multirow{3}{*}{\begin{tabular}[c]{@{}c@{}}Offset\\ No.\end{tabular}} & \multirow{3}{*}{\begin{tabular}[c]{@{}c@{}}Cumulative \\ Exposure \\ Time (s)\end{tabular}} & \multicolumn{4}{c|}{\textbf{Supernova}}                                                                                                                               \\ \cline{1-3} \cline{6-9} 
\multicolumn{1}{|c|}{\multirow{2}{*}{\textit{ID}}} & \multicolumn{2}{c|}{\textit{Coordinates (J2000)}}                              &                                                                       &                                                                                                & \multicolumn{1}{c|}{\multirow{2}{*}{\textit{ID}}} & \multicolumn{2}{c|}{\textit{Coordinates (J2000)}}                                & \multirow{2}{*}{\textit{Type}} \\ \cline{2-3} \cline{7-8}
\multicolumn{1}{|c|}{}                             & \multicolumn{1}{c|}{\textit{RA (Deg)}}           & \textit{DEC (Deg)}          &                                                                       &                                                                                                & \multicolumn{1}{c|}{}                             & \multicolumn{1}{c|}{\textit{RA (Deg)}} & \multicolumn{1}{c|}{\textit{DEC (Deg)}} &                                \\ \hline
\multicolumn{1}{|c|}{\multirow{2}{*}{NGC-1404}}    & \multicolumn{1}{c|}{\multirow{2}{*}{54.716333}}  & \multirow{2}{*}{-35.594389} & \multirow{2}{*}{2}                                                    & \multirow{2}{*}{2000}                                                                          & \multicolumn{1}{c|}{2007on}                       & \multicolumn{1}{c|}{54.712496}         & \multicolumn{1}{c|}{-35.575311}         & Ia                             \\ \cline{6-9} 
\multicolumn{1}{|c|}{}                             & \multicolumn{1}{c|}{}                            &                             &                                                                       &                                                                                                & \multicolumn{1}{c|}{2011iv}                       & \multicolumn{1}{c|}{54.713958}         & \multicolumn{1}{c|}{-35.592222}         & Ia                             \\ \hline
\multicolumn{1}{|c|}{NGC-3244}                     & \multicolumn{1}{c|}{156.370167}                  & -39.827556                  &    3                                                                   &     5040                                                                                           & \multicolumn{1}{c|}{2010ev}                       & \multicolumn{1}{c|}{156.370792}        & \multicolumn{1}{c|}{-39.830889}         & Ia                             \\ \hline
\multicolumn{1}{|c|}{NGC-3351}                     & \multicolumn{1}{c|}{160.990417}                  & 11.703806                   &    2                                                                   &     1240                                                                                           & \multicolumn{1}{c|}{2012aw}                       & \multicolumn{1}{c|}{160.974000}        & \multicolumn{1}{c|}{11.671639}          & II                             \\ \hline
\multicolumn{1}{|c|}{NGC-4424}                     & \multicolumn{1}{c|}{186.79838}                   & 9.42067                     &    4                                                                   &     8140                                                                                           & \multicolumn{1}{c|}{2012cg}                       & \multicolumn{1}{c|}{186.803458}        & \multicolumn{1}{c|}{9.420333}           & Ia                             \\ \hline
\multicolumn{1}{|c|}{\multirow{2}{*}{NGC-6962}}    & \multicolumn{1}{c|}{\multirow{2}{*}{311.829434}} & \multirow{2}{*}{0.320801}   &    \multirow{2}{*}{2}                                                                  &     \multirow{2}{*}{5000}                                                    & \multicolumn{1}{c|}{2002ha}                       & \multicolumn{1}{c|}{311.827417}        & \multicolumn{1}{c|}{0.312667}           & Ia                             \\ \cline{6-9} 
\multicolumn{1}{|c|}{}                             & \multicolumn{1}{c|}{}                            &                             &                                                                       &                                                                                                & \multicolumn{1}{c|}{2003dt}                       & \multicolumn{1}{c|}{311.823167}        & \multicolumn{1}{c|}{0.311889}           & Ia                             \\ \hline
\multicolumn{1}{|c|}{NGC-7721}                     & \multicolumn{1}{c|}{354.702708}                  & -6.517861                   &    2                                                                   &     4000                                                                                           & \multicolumn{1}{c|}{2007le}                       & \multicolumn{1}{c|}{354.701708}        & \multicolumn{1}{c|}{-6.522583}          & Ia                             \\ \hline
\multicolumn{1}{|c|}{NGC-7780}                     & \multicolumn{1}{c|}{358.384042}                  & 8.118139                    &    2                                                                   &     2000                                                                                           & \multicolumn{1}{c|}{2001da}                       & \multicolumn{1}{c|}{358.386583}        & \multicolumn{1}{c|}{8.117389}           & Ia                             \\ \hline
\multicolumn{1}{|c|}{UGC-11723}                    & \multicolumn{1}{c|}{320.072833}                  & -1.684333                   &    2                                                                   &     960                                                                                           & \multicolumn{1}{c|}{2006cm}                       & \multicolumn{1}{c|}{320.07275}         & \multicolumn{1}{c|}{-1.684083}          & Ia                             \\ \hline
\multicolumn{1}{|c|}{UGC-12859}                    & \multicolumn{1}{c|}{359.214792}                  & 5.508417                    &    2                                                                   &     2000                                                                                           & \multicolumn{1}{c|}{2007fb}                       & \multicolumn{1}{c|}{359.218208}        & \multicolumn{1}{c|}{5.508833}           & Ia                             \\ \hline
\end{tabular}
\caption{List of galaxies analysed in this work and of SNe hosted within them.}
\label{tab:hgid}
\end{table*}

FORS2 is a dual-beam polarimeter that separates the incoming light flux into two beams with perpendicular polarization states named the ordinary ($O$) and extra-ordinary ($E$) beams. It does this by combining a half-waveplate (HWP), which induces a phase shift of $\pi$ between the two perpendicular light components, and a Wollaston prism (WP) which then diverges them. The two components can be measured at different position angles $\theta_k$ of the optical axis of the HWP. As discussed in \cite{patatrom}, the same field is observed at least at four HWP positions chosen at constant intervals of $\Delta\theta = \pi / 8$ to minimize errors.\par

In this work, we obtain the normalized Stokes parameters ($q \equiv Q / I$ and $u \equiv U / I$) from the $O$ and $E$ beam fluxes, $f_{O, k}$ and $f_{E, k}$, following the "ratio" method presented by \cite{bagnulo}. This method is designed to cancel out any spatial polarization effects that may arise due to small WP anisotropies. For HWP optical axis position angle intervals of $\Delta\theta = \pi / 8$,  the parameters are given by

\begin{align}\label{eq:QandUbag}
    q = \frac{\sqrt{R_{0,2}} - 1}{\sqrt{R_{0,2}} + 1} \:, \nonumber \\
    u = \frac{\sqrt{R_{1,3}} - 1}{\sqrt{R_{1,3}} + 1} \:,
\end{align}

\noindent where each $R_{k,k+2}$, with $k \in \{0,1\}$, is the ratio of flux ratios at position angles $\theta_k$ and $\theta_{k+2}$, given by

\begin{align}\label{eq:Ri}
    R_{k,k+2} = \frac{f_{O,k} f_{E,k+2}}{f_{E,k} f_{O,k+2}} \: .
\end{align}

\noindent The polarization degree, $p$, and the angle, $\chi$, are then obtained using

\begin{align}\label{eq:PandX}
    p \:&=\: \sqrt{q^2+u^2} \nonumber \\
    \chi \:&=\: \frac{1}{2}\arctan\Big(\frac{u}{q}\Big)\:.
\end{align}

\noindent The polarization degree extracted from the previous relation is biased; therefore, we instead use the unbiased estimator $\hat{p}$ introduced by \cite{debias}, given by 

\begin{align}
    \hat{p} \:&=\: p - b^2 \frac{1 - \exp{\big[-\frac{p^2}{b^2}\big]}}{2p} \:, 
\end{align}

\noindent where $b^2$ is

\begin{align}\label{eq:pbias}
    b^2 \:=\: \frac{q^2\: s^2_u + u^2\: s^2_q}{q^2 + u^2} \:. 
\end{align}

\noindent In the remainder of this text, when referring to measurements of the polarization degree, $p \equiv \hat{p}$.\par

Dual-beam polarimetry of extended objects requires an extra step. To avoid the superposition of the O and E beams in the detector, a polarization strip mask is applied to alternatively filter out one of the two components. This mask leads to the loss of half of the spatial field, which then prompts the need for complementary observations with an offset relative to the initial position. For more details on dual-beam polarimetry of extended objects, the reader is referred to \citep{patatrom, bagnulo, tips}.\par

Our data include observations at 4 HWP position angles and at least two offsets for each galaxy. A series of scripts, written in R\footnote{\href{https://www.r-project.org/}{https://www.r-project.org/}}, were developed to process the data. These scripts are publicly available\footnote{\href{https://github.com/p41nk1ll3r/FORS2-Photo-Polarimetry}{Calibration and Reduction Scripts}} and perform the following tasks: 

\begin{itemize}
    \item create master bias files for each CCD chip;
    \item create master flat files for each CCD chip;
    \item create calibrated and joined (combining chips) acquisition, standard, and science files;
    \item split the O beams and E beams (in a single file) into different files;
    \item merge O beams (of different files for the same target) into a single file (and the same for E beams);
    \item remove cosmic rays;
    \item estimate the background;
    \item perform photometry on field stars matched to Gaia \citep{gaia3} to estimate the ISP of the MW for that field;
    \item bin the O beam and E beam images, into larger bins to increase the signal-to-noise ratio;
    \item calculate normalized Stokes parameters $q \equiv \frac{Q}{I}$ and $u \equiv \frac{U}{I}$ images corrected for background, MW ISP, and instrumental polarization, as well as polarization degree, $p$, and angle $\chi$ image files.
\end{itemize}

Detailed information on the methodology and implementation of these scripts is described in the companion paper \citep{self-pipe}. The results obtained are analyzed in the following section. 

\section{Dust scattering and absorption}\label{sec:models}

\begin{figure*}
    \centering
    \begin{subfigure}{.35\textwidth}
        \centering
        \includegraphics[width =\textwidth]{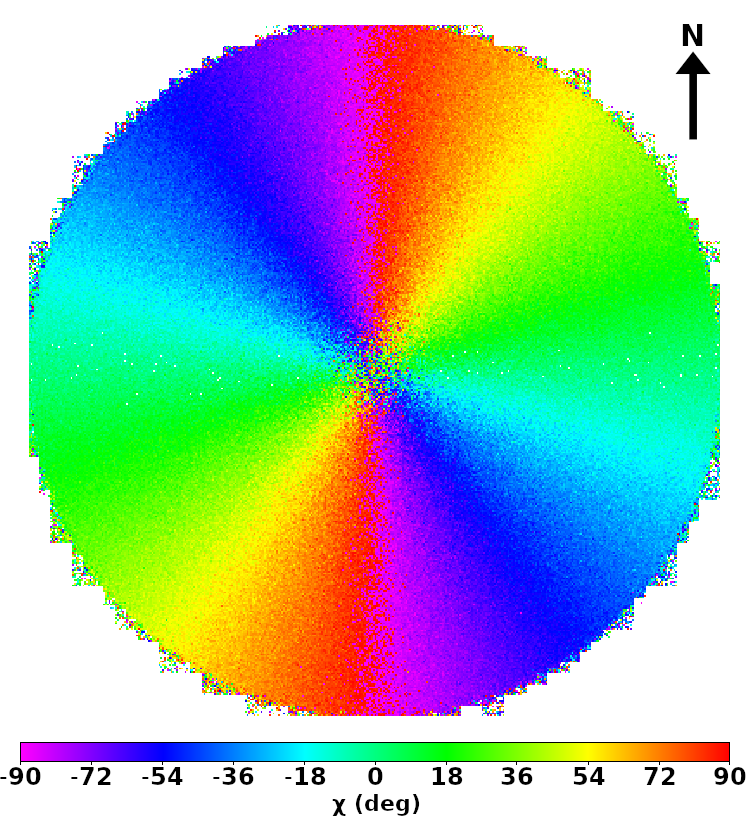}
        \caption{$\tau = 1$}
        \label{fig:scatAnga}
    \end{subfigure}
    \hspace{5mm}
    \begin{subfigure}{.35\textwidth}
        \centering
        \includegraphics[width =\textwidth]{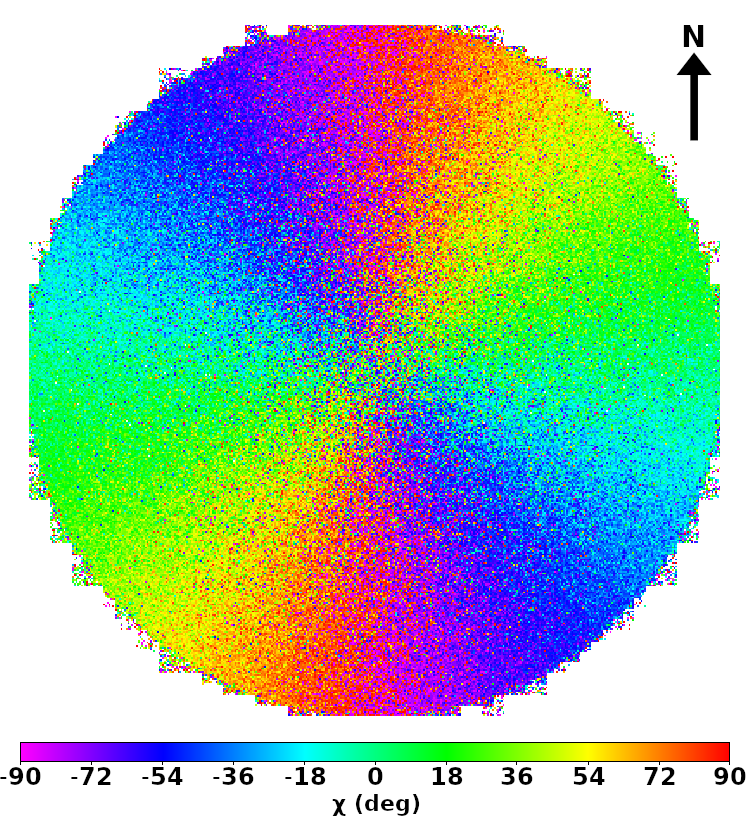}
        \caption{$\tau = 5$}
        \label{fig:scatAngb}
    \end{subfigure}
    \caption{Polarization angle in degrees, measured relative to the North, maps at $\lambda = $ \SI{228.5}{nm} for two simulations of an emitting spheroidal source inside an isotropic, homogeneous spherical dust medium. Detailed descriptions of the models can be found in Appendix \ref{app:A}. The optical depth $\tau$, of the medium of the simulation on the right is 5$\times$ higher than that of the one on the left, which results in more light being scattered multiple times before reaching the LOS, leading to randomization of the polarization angle pattern and depolarization.}
    \label{fig:scatAng}
\end{figure*}

\begin{figure*}
    \centering
    \includegraphics[width =.85\textwidth]{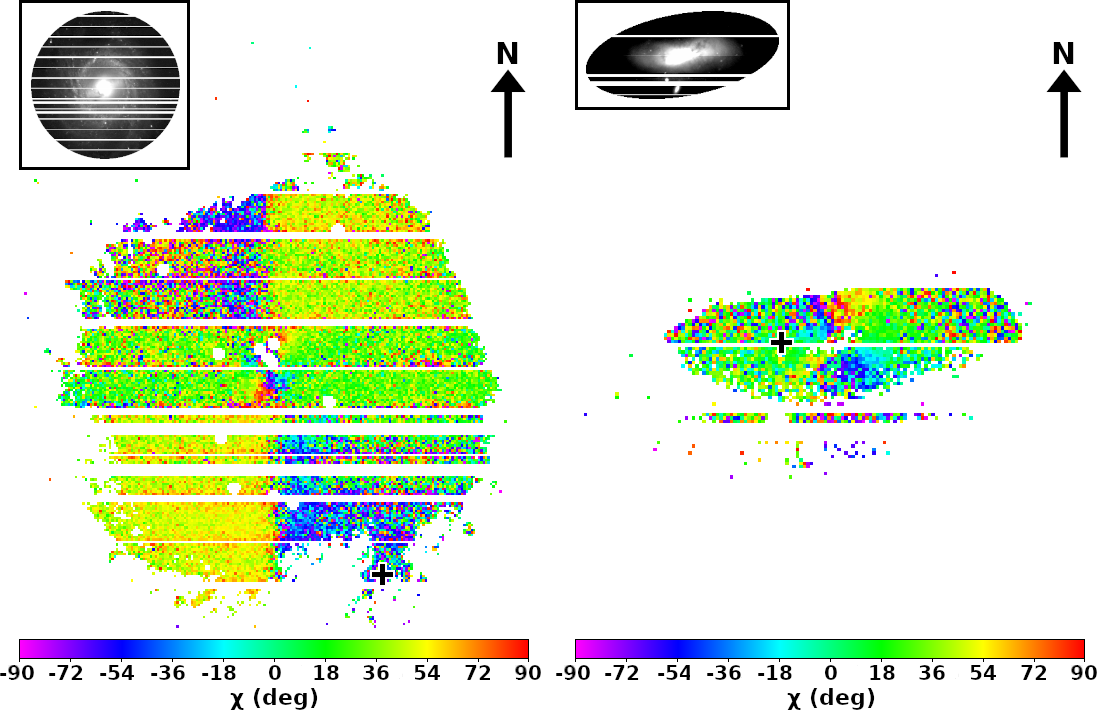}
    \caption{Maps of the polarization angle in degrees, measured relative to the North, with $B$ filter. Left: NGC-3351 with SN 2012aw approximate location marked with a black cross; Right: NGC-4424 with SN 2012cg approximate location marked with a black cross. The patterns displayed are compatible with polarization due to the scattering of light of a dominating central source inside a dusty medium. Milky Way stars are masked by white circles. A flux image of the galaxies, using the same filter, is presented at the top left corner of each map.}
    \label{fig:scatGals}
\end{figure*}

In this section we discuss the general expected polarization patterns from the two main processes of dust scattering and dust absorption. In the first case, as registered in observations \citep{scarr2} and Monte Carlo Radiative Transfer (MCRT) simulations \citep{peest}, for a dominant light source at the center of a dust medium, the polarization angle of light scattered a single time into the LOS is expected to be perpendicular to the direction of emission, thus displaying a circular axisymmetric pattern around the light source(s): Fig. \ref{fig:scatAnga} illustrates this situation for two simple models simulated using the radiative transfer code \texttt{SKIRT} \citep[][more information about these models is provided in Appendix \ref{app:A}]{skirt1,skirt2}. In optically thick media, this pattern may erode as a result of multiple scattering into random orientations and the consequent depolarization. Due to the lower flux levels (less light scattered into the LOS) and lower polarized fraction, randomization of the polarization angle is bound to be more spatially frequent, as can be seen in Fig. \ref{fig:scatAngb}. Likewise, if there are other comparably luminous light sources in the field, the pattern will erode due to the mixing of light from different sources. Fig. \ref{fig:scatGals} provides exemplary polarization angle maps from our data set whose patterns are compatible with polarization mostly due to scattering.\par 

\begin{figure}
    \includegraphics[width =.49\textwidth]{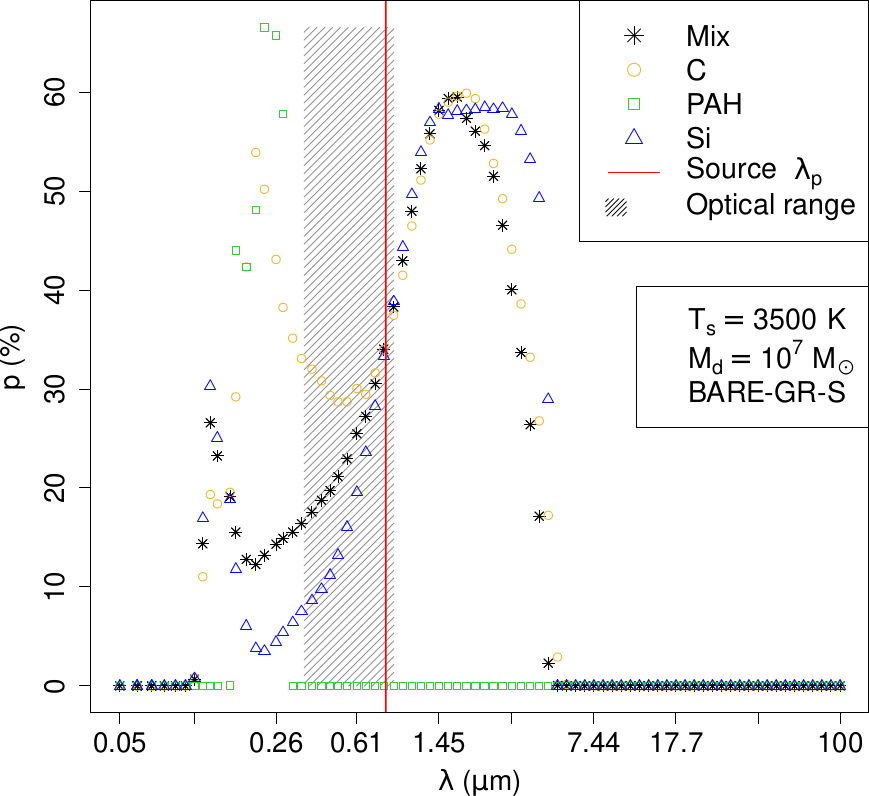}
    \includegraphics[width =.49\textwidth]{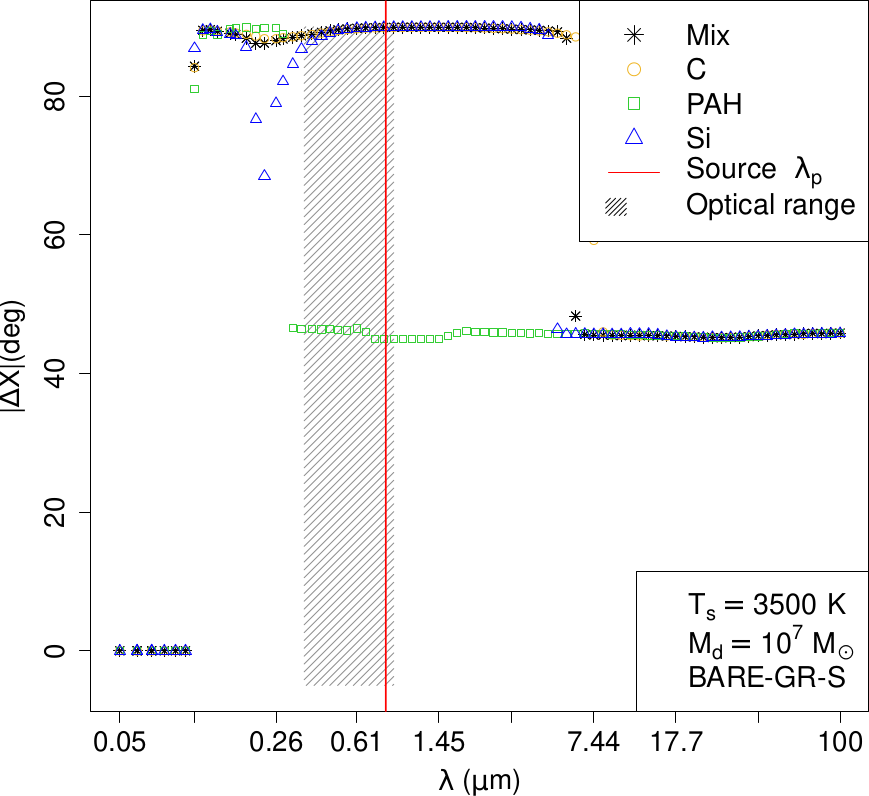}
    \caption{Scattering wavelength dependence of the spatial average of polarization degree (upper) and polarization angle relative to the radial direction (lower) from radiative transfer simulations of a bright source with $T_s = \SI{3500}{\K}$ located at the center of dusty media composed of: silicates (blue triangles), graphites (yellow circles), Polycyclic Aromatic Hydrocarbons (green squares) and the \texttt{BARE-GR-S} dust mixture of the those three compounds \citep[][black stars]{zubko}. The shaded area indicates the optical wavelength range and the red line the emitting sources peak wavelength, $\lambda_p = \SI{0.828}{\um}$.}
    \label{fig:scat-polwave}
\end{figure}

It is also important to mention that the wavelength dependence of the polarization degree for pure scattering is not universal and will follow different trends according to the dust characteristics, as is shown in the MCRT simulation of Figure~\ref{fig:scat-polwave}. There is no single empirical law for scattering to model the observational data.  Interestingly, in the case of a typical MW dust mix, the polarization reaches a minimum within the optical region ($\sim0.61 \mu m$) which could be observed with our data. Furthermore, the wavelength dependence of the polarization angle is expected to be mostly constant in the optical region for pure scattering as seen in the lower Fig.~\ref{fig:scat-polwave}.\par

In the second case of absorption by aligned dust grains, for optical and NIR radiation, the polarization angle of directly illuminated regions will behave differently. A medium composed of spheroidal dust grains embedded in an external magnetic field is expected to have those grains aligned (even if only partially) with their shorter axis parallel to that field \citep[][they may rotate around that parallel component]{Su2013}; this will lead to the absorption of UV/optical light preferentially for light oscillating parallel to the longer axis (perpendicular to the magnetic field), which in turn will result in the polarization of the detected light to be parallel to the magnetic field. Conversely, the absorbed energy will then be re-emitted in MIR/FIR wavelengths with polarization perpendicular to the magnetic field. We would also expect a rather constant polarization angle as a function of wavelength for our purely optical $BVRI$ observations. In fact, in near edge-on galaxies, the optical polarization angle has been observed to be parallel to the disk \citep{scarr2, montgomery}, while in face-on galaxies it appears to follow the direction tangential to the arms \citep{scarr}. Definite predictions are still difficult to make due to a general lack of knowledge regarding the galactic magnetic field structure at both large (dynamo effect) and local scales (flow of charged particles); and current radiative transfer codes lack the capability to model magnetic field and dust alignment. Moreover, for the face-on case similar polarization orientation patterns may arise through both absorption and scattering processes \citep{scarr, peest}, which will then require a spatially comprehensive wavelength dependence analysis of both the polarization degree and angle to attempt differentiation. 

In fact, as mentioned earlier, the dependence of the polarization degree $p$ with wavelength $\lambda$ for dust alignment generally follows an empirical Serkowski law \citep{serkowski}:

\begin{equation}\label{eq:serk}
    p(\lambda) = p_{max} \exp\bigg[-k_p \ln^2{\bigg(\frac{\lambda}{\lambda_{max}}\bigg)}\bigg] \:,    
\end{equation}

\noindent where $p_{max}$ is the maximum polarization degree, $\lambda_{max}$ is the wavelength at which $p_{max}$ occurs, and $k_p$ gives the width of the curve and has been empirically connected to $\lambda_{max}$ \citep{wilking} through: 

\begin{align}\label{eq:wilk}
    k_p = -0.1 + 1.86\,\lambda_{max}  \:.
\end{align}

\noindent${\lambda_{max}}$ is acknowledged to be connected to the size of dust grains in the medium \citep{draine} and has been shown to be linearly correlated with the total-to-selective extinction ratio or "reddening law" $R_V$ \citep{rv, vrba}, which expresses the interstellar dust-induced starlight reddening and is defined as:

\begin{equation}\label{eq:RvRed}
    R_V = \frac{A_V}{E_{B-V}}\:\:,
\end{equation} 

\noindent where $A_V$ is the total extinction in the V band and $E_{B-V}$ is the color excess, or selective extinction ($E_{B-V} = A_B - A_V$), of the V band relative to the B band. The empirical relation of $R_V$ with maximum wavelength of polarization is \citep{rv, vrba}:
\begin{equation}\label{eq:Rv}
    R_V \approx 5.5\,\lambda_{max}\:. 
\end{equation}

 The relation expressed by Eq. \ref{eq:serk} should appear in regions with little contribution from light of surrounding sources scattered into the LOS. Its wavelength dependence is clearly different from the scattering case (see Fig.~\ref{fig:scat-polwave}), offering a means to disentangle both effects. Some example Serkowski laws fitted to our data are shown in Fig.~\ref{fig:exSerk}. However, when multiple dust clouds are present in the LOS and are composed of different dust properties, a simple Serkowski relation may not be followed and extra care must be taken to model these cases \citep{mandarakas24}. More generally, both scattering and absorption may contribute to a polarized signal, resulting in unexpected polarization degree and angle dependence on wavelength. Separating their contributions is not a trivial task. \par

\section{Analysis}\label{sec:meth}
For each galaxy, we visually inspect the 2D maps of the polarization degree $p$, and angle $\chi$, searching for clear signatures of polarization due to scattering into the LOS or absorption by aligned dust grains. We also look at the relationship between the polarization degree, $p$, and the effective wavelength, $\lambda_c$, of the measurement band filters, for different regions of the galaxies, both those whose polarization is dominated by scattering and those without scattering signatures in their polarization maps. We start by analyzing bins of $5\times5$ \SI{}{pix^2} with the projected length of these bins spanning from \SI{0.06}{kpc} to \SI{.45}{kpc}, depending on the galaxy; then we repeat the analysis with bins of $25\times25$ \SI{}{pix^2}. In each case, if the bin that includes the coordinates of the SN was not successfully fitted with a Serkowski law, our $R_V$ estimate will be based on the average of neighbors at a Manhattan distance of 1 bin. \par

\begin{figure}
    \centering
    \includegraphics[width=\linewidth]{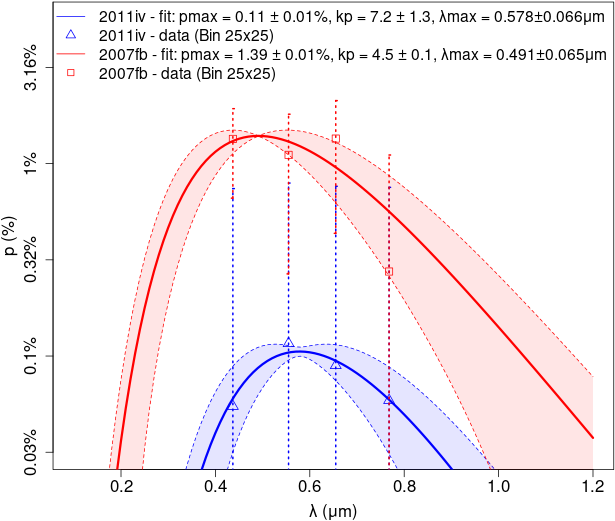}
    \caption{Two different Serkowski relations, displaying how each parameter controls the curve. $p_{max}$ yields the maximum polarization degree and occurs at $\lambda = \lambda_{max}$, while $k_p$ relates to the width of the curve. The blue and red lines represent Serkowski models for the vicinity of SNe 2011iv and 2007fb, respectively, while the same colored symbols indicate the polarimetry that yielded those models. The shaded areas indicate the 1$\sigma$ confidence interval of the models predictions.}
    \label{fig:exSerk}
\end{figure}

For every instance mentioned above, we perform robust fitting of a Serkowski relation. This was achieved by first sampling regions $p(\lambda) \pm \sigma(p(\lambda))$ estimates at each band, 10000 times, with each point in the sample being assigned a weight, $w = 1 / \sigma^2(p(\lambda))$. The nonlinear least squares algorithm \texttt{'port'} within the function \texttt{nls} \citep{nls1, nls2} of the \texttt{stats} library was used. This algorithm requires the definition of initial values and bounds for the model parameters. The initial values for the Serkowski parameters were set according to the data: $p_{max}$ as the maximum of $p$ measured across the four wavebands, $\lambda_{max}$ as the $\lambda_c$ of the waveband that holds the maximum of $p$ within the sample, and $k_p = -0.1 + 1.86 \,\lambda_{max}$ (following  \citep{wilking}). The following bounds were also defined: $0.000001 \leq p_{max} \leq 1$ and $\lambda_{max} \geq 0.01$ \SI{}{\um}. In Fig. \ref{fig:exSerk} we plot two Serkowski relations, with different parameters, obtained throughout this analysis for the galaxy positions around SN 2011iv and 2007fb.\par

With the fits in different regions of the galaxies, we were able to estimate $R_V$ in the projected vicinity of the hosted SNe. Since our estimates rely on the Serkowski relation given by Eq. \ref{eq:Rv}, and scattering effects could obfuscate our estimates, we employed a set of criteria to select the parameters that best fit a pure Serkowski relation, namely: 

\begin{itemize}
    \item $\lambda_{max}$ must be within the range of the observations, which in our case is $0.388 \leq \lambda_{max} \leq 0.870$ \SI{}{\um} (which is equivalent to $2.13 \leq R_V \leq 4.79$);
    \item $\lambda_{max} / \sigma(\lambda_{max}) > 5$;
    \item $k_p > 0$;
    \item the correlation between Serkowski model predictions, $\tilde{p}$, and the measurements should be larger than 0.75: $corr\big(\tilde{p}(\lambda_c), p(\lambda_c)\big) > 0.75$;
    \item the mean normalized absolute residuals between best model and the data should be less than 25\%: 
    \begin{align}
        res(\%) = E\Big(\frac{|\tilde{p}(\lambda_c) - p(\lambda_c)|}{p(\lambda_c)}\Big)\times100\% < 25\%\:.
    \end{align}
\end{itemize}

\noindent The first two criteria are soft filters, meaning that a fit not complying with these conditions is still saved for later inspection and used for spatial inferences. The last three criteria ensure that the polarization follows a Serkowski relation.  When estimates for the SN vicinity are  available for both bin sizes, the estimate based on fewer pixels is favored since it should be more representative of the SNe vicinity. The estimate for the location of SN 2001da was obtained from the $5\times5$ pix$^2$ bin that includes those SN coordinates, while for SN 2007le the mean of two neighboring $5\times5$ pix$^2$ bins. For SNe 2006cm, 2007fb, and 2011iv the estimates come from the $25\times25$ pix$^2$ bins of their coordinates. The results are presented in Tab.~\ref{tab:sne_Rv_vanila}.\par

\begin{figure*}
    \centering
    \includegraphics[width =.9\textwidth]{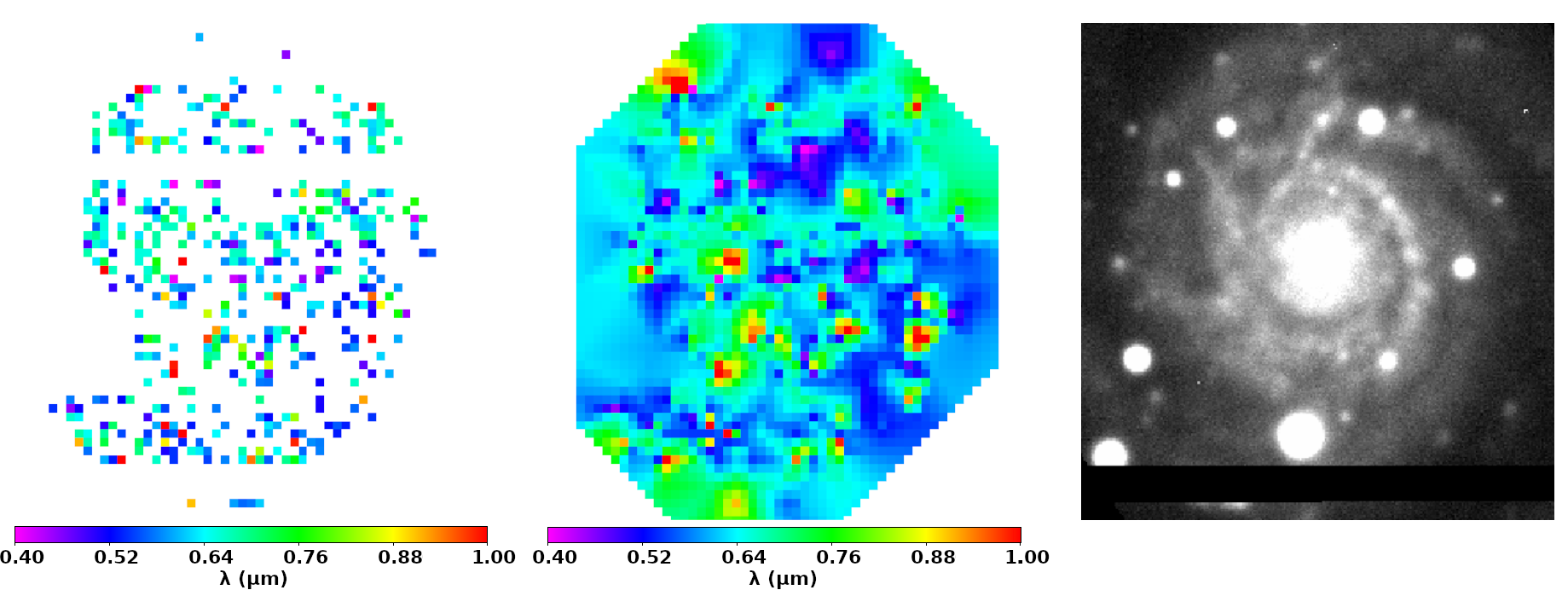}
    \caption{Left: $\lambda_{max}$ map for NGC-3244, obtained by fitting Serkwoski laws to each $5\times5$ \SI{}{pix^2} bin and filtering the results of those fits using the criteria defined in Section \ref{sec:meth}. Center: $\lambda_{max}$ map inference performed, by \texttt{INLA}, on the $\lambda_{max}$ map on the left. Right: \texttt{b\_HIGH} pointing image of NGC-3244 (MW field stars regions were masked during the data reduction steps). All images are presented at the same spatial scale.}
    \label{fig:inlaRv}
\end{figure*}

To obtain an $R_V$ estimate for the remaining SNe locations (2002ha, 2003dt, 2007le, 2007on, 2010ev, 2012aw and 2012cg), for which no bin or neighboring bins provided an accurate Serkowski estimate, we employed \texttt{INLA} \citep{inla, inlaR} to spatially infer from sparse bin maps of $\lambda_{max}$ and $R_V$ the values at the SN posiions. \texttt{INLA} is an approximate Bayesian inference method that accounts for spatial correlations between observed data points to not only predict spatial intermediate unobserved points but also correct noisy observed ones while associating a variance to its predictions. It has already been shown that \texttt{INLA} is capable of recovering structures in scalar and vector fields with great fidelity out of sparse sets of observations being robust to noise injections \citep{inlaApp1,emulart,majda}. With \texttt{INLA} it is also possible to infer structures never seen before \citep{inlaApp2}. In Fig. \ref{fig:inlaRv}, we provide an example of spatial inference performed by \texttt{INLA} on a map of $\lambda_{max}$. For a more detailed comparison between the $R_V$ values inferred from Serkowski fits and INLA reconstructions, and the choice of the best $R_V$ for each SN position, we refer the reader to Appendix~\ref{app:Rv}.\par

\begin{table*}[ht]
\centering
{\renewcommand{\arraystretch}{1.4}
\begin{tabular}{|c|c|c|c|c|c|c|}
\hline
SN & Host Galaxy & $\lambda_{max}$ (\SI{}{\um}) & $R_V$ & Method & Projected Area (kpc$^2$)\\ \hline
2007on & NGC-1404 & - & $3.4\pm1.0$ & \texttt{INLA} & $0.12 \times 0.12$ \\ \hline
2011iv & NGC-1404 & $0.578\pm0.066$ & $3.18\pm0.36$ & Fit of $25\times25$ Bin & $0.58 \times 0.58$ \\ \hline
2010ev & NGC-3244 & - & $3.16\pm0.65$ & \texttt{INLA} & $0.17 \times 0.17$ \\ \hline
2012aw & NGC-3351 & - & $3.33\pm0.75$ & \texttt{INLA} & $0.06 \times 0.06$ \\ \hline
2012cg & NGC-4424 & - & $3.50\pm0.82$ & \texttt{INLA} & $0.09 \times 0.09$ \\ \hline
2002ha & NGC-6962 & - & $3.04\pm1.28$ & \texttt{INLA} & $0.34 \times 0.34$ \\ \hline
2007le & NGC-7721 & $0.58\pm0.12$ & $3.17\pm0.65$ & Mean fit of two neighboring $5\times5$ bins & $0.38 \times 0.38$ \\ \hline
2001da & NGC-7780 & $0.723\pm0.065$ & $3.98\pm0.36$ & Fit of $5\times5$ Bin & $0.45 \times 0.45$ \\ \hline
2006cm & UGC-11723 & $0.476\pm0.065$ \dag & $2.62\pm0.36$ \dag & Fit of $25\times25$ Bin & $2.17 \times 2.17$ \\ \hline
2007fb & UGC-12859 & $0.491\pm0.065$ & $2.70\pm0.36$ & Fit of $25\times25$ Bin & $2.27 \times 2.27$ \\ \hline
\end{tabular}
}
\caption{Estimates of $\lambda_{max}$ and $R_V$ in the vicinity of the SNe. These estimates were obtained either via robust fitting of the Serkowski relation to different bins and bin selections of the multi-waveband $p$ maps obtained using the reduction procedures outlined in Section \ref{sec:data}, or by spatial inference with \texttt{INLA} on maps of fitted estimates of $R_V$. The projected area, in \SI{}{kpc}$^2$, corresponding to the bin selection is also provided. Inferred estimates always correspond to a $5\times5$ bin. The estimate marked with (\dag) is based on a Serkowski model fit with only three bands.}
\label{tab:sne_Rv_vanila}
\end{table*}

\section{Results and Discussion}\label{sec:disc}
\subsection{Comparison of $R_V$ estimates from photopolarimetry and from SN light curves}
The $R_V$ values estimated using photopolarimetry are assigned to a median projected area of $0.022\pm0.017$ \SI{}{kpc^2}, which corresponds to a squared vicinity with a length of $\sim$ \SI{150}{pc}, this is much larger than the usual scale of an \ion{H}{II} region ($\sim$\SI{30}{pc}) which means that the properties inferred for these bins may likely be averages of different dust structures. The obtained $R_V$ values range from $3.01\pm0.54$ to $3.98\pm0.36$ with an average of $3.30\pm0.74$, which is consistent with the average $R_V \approx 3.1$ extinction law of the Milky Way \citep{CCM, Odonnell94} and distinct from $R_V \approx 1.7$ which is associated with very reddened Type Ia SNe \citep{carnagie}.\par

\begin{figure}
    \centering
    \includegraphics[width =\linewidth]{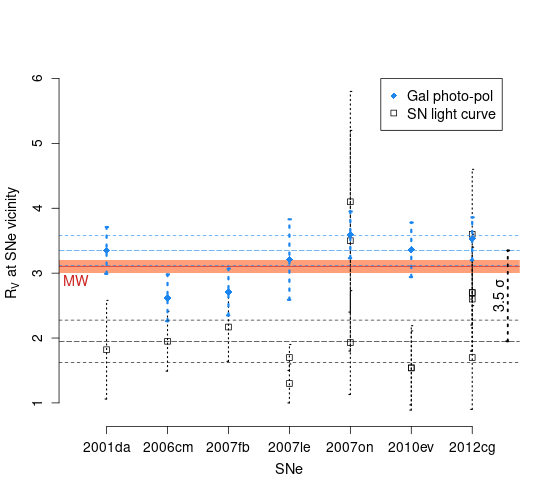}
    \caption{Comparison of $R_V$ estimated at the SN vicinity with estimates from SN observations. Blue dots are estimates obtained using late-time (several years post maximum light) multi-waveband photopolarimetry at the vicinity of SNe; black squares are estimates based on model fitting to color light curve data of those SNe (\citep{Mandel11, Phillips13, carnagie, Amanullah2015, claudia16}). The median, and its uncertainty, of each estimation are presented in their corresponding color. The fiducial $R_V$ for the Milky Way is presented by a red line with its uncertainty being the shaded orange area. A $3.5\sigma$ separation is observed between the median of each kind of estimate.}
    \label{fig:Sne_oldxnew}
\end{figure}

As shown in Fig. \ref{fig:Sne_oldxnew} and Tab. \ref{tab:Rvs_vs_lit}, the $R_V$ obtained from photopolarimetry differ by up to 3.3$\sigma$ from estimates derived from SN light curve analysis \citep{Mandel11, Phillips13, carnagie, Amanullah2015, claudia16}. The values obtained by \cite{Amanullah2015} at different epochs of SN 2012cg hint at $R_V$ varying with the time after peak luminosity (in the \texttt{B} band). This had already been shown to be the case for the sample of SNe studied by \cite{forster13}, with two scenarios proposed: multiple scattering processes due to the presence of CSM \citep{Amanullah2011}; or, the progressive destruction of sufficiently nearby material \citep{forster13}, which would lead to a change in reddening over time. Moreover, analysis by \cite{bulla18} of photometric data of 48 SNe showed that 15 of those had a time variable $E_{B-V}$ and they predicted that the dust responsible for that variation was within the CSM range for 4 SNe and as part of nearby patchy ISM for the remaining 11. Our photopolarimetric data lack the spatial resolution necessary to make any assessments regarding the time variability of the SNe CSM but it is worth noting that not a single area within our galaxies is consistent with such small $R_V$ values as found for these peculiar SNe~Ia.\par

\begin{table*}
\centering
{\renewcommand{\arraystretch}{1.4} 
\begin{tabular}{|c|c|c|l|c|}
\hline
Supernova                 & $R_V$                   & Reference / Method                       & This Work                       & $\Delta R_V(\sigma)$ \\ \hline
2001da                  & $1.82\pm^{0.76}_{0.53}$ &  \cite{Phillips13} / (a)     & $3.98\pm0.36$                   & 2.57           \\ \hline
2006cm                  & $1.95\pm^{0.46}_{0.33}$ &  \cite{Phillips13} / (a)     & $3.01\pm0.54$*                   & 1.49          \\ \hline
2007fb                  & $2.17\pm^{0.53}_{0.60}$ &  \cite{Phillips13} / (a)     & $3.15\pm0.58$*                   & 1.25           \\ \hline
\multirow{2}{*}{2007le} & $1.3\pm^{0.3}_{0.2}$    &  \cite{carnagie} / (b)       & \multirow{2}{*}{$3.28\pm0.51$*} & 3.34           \\ \cline{2-3} \cline{5-5} 
                          & $1.7\pm^{0.2}_{0.2}$    &  \cite{carnagie} / (c)       &                                 & 2.88           \\ \hline
\multirow{3}{*}{2007on} & $4.1\pm^{2.0}_{1.7}$    &  \cite{carnagie} / (b)       & \multirow{3}{*}{$3.4\pm1.0$*}  & 0.35           \\ \cline{2-3} \cline{5-5} 
                          & $3.5\pm^{1.7}_{1.4}$    &  \cite{carnagie} / (c)       &                                 & 0.05           \\ \cline{2-3} \cline{5-5} 
                          & $1.93\pm^{0.80}_{0.61}$ &  \cite{Phillips13} / (a)     &                                 & 1.15           \\ \hline
\multirow{2}{*}{2010ev} & $1.54\pm0.65$           &  \cite{claudia16} / (b)      & \multirow{2}{*}{$3.16\pm0.65$}  & 1.76           \\ \cline{2-3} \cline{5-5} 
                          & $1.54\pm^{0.57}_{0.59}$ &  \citep{Phillips13} / (a)     &                                 & 1.87           \\ \hline
\multirow{5}{*}{2012cg} & $2.7\pm0.5$             &  \cite{Mandel11} / (a)        & \multirow{5}{*}{$3.50\pm0.82$*}  & 0.83           \\ \cline{2-3} \cline{5-5} 
                          & $2.6\pm^{0.8}_{0.7}$    &  \cite{Amanullah2015} / (b)  &                                 & 0.78           \\ \cline{2-3} \cline{5-5} 
                          & $2.7\pm^{0.9}_{0.7}$    &  \cite{Amanullah2015} / (c)   &                                 & 0.66           \\ \cline{2-3} \cline{5-5} 
                          & $1.7\pm^{0.8}_{0.6}$    &  \cite{Amanullah2015} / (c')  &                                 & 1.57           \\ \cline{2-3} \cline{5-5} 
                          & $3.6\pm^{1.5}_{1.0}$    &  \cite{Amanullah2015} / (c'') &                                 & 0.08           \\ \hline
\end{tabular}
}
\caption{Comparison between published $R_V$ directly from SNe data and our estimates at the SN vicinity in the host. Methods: a) Fit of SNe Colors at peak with MCMC; b) Fit of SNe Colors at peak following  \citep{Odonnell94}; c) Fit of SNe Colors at peak following  \citep{F99}; c') same as (c) but on [-5;+5] days from maximum sub-sample; c'') same as (c) but on [+10;+20] days after maximum sub-sample. In the 4th column: $R_V$ is estimated in this work by fitting the Serkowski relation to photopolarimetry of the host galaxy at the SN vicinity; (*) indicates the use of \texttt{INLA} for spatial inference using the values obtained through the previous method.}
\label{tab:Rvs_vs_lit}
\end{table*}

\begin{figure}
    \centering
    \includegraphics[width =\linewidth]{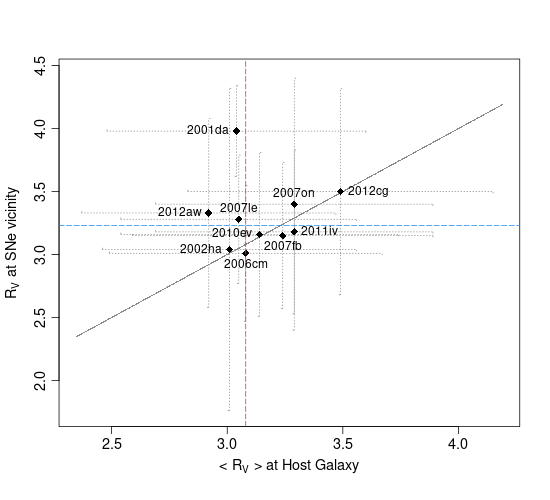}
    \caption{$R_V$ at the vicinity of the hosted SNe \textit{versus} the global $R_V$ for the present sample of SN hosts. The global $R_V$ is a median estimated using all bins well modeled by a Serkowski relation.  The red dashed line marks the median of the host galaxies global $R_V$, while the blue dashed line presents the median of the $R_V$ at the SNe vicinity. The black line represents an ideal 1:1 relation.}
    \label{fig:Rv_SNxHG}
\end{figure}

\begin{figure}
    \centering
    \includegraphics[width=\linewidth]{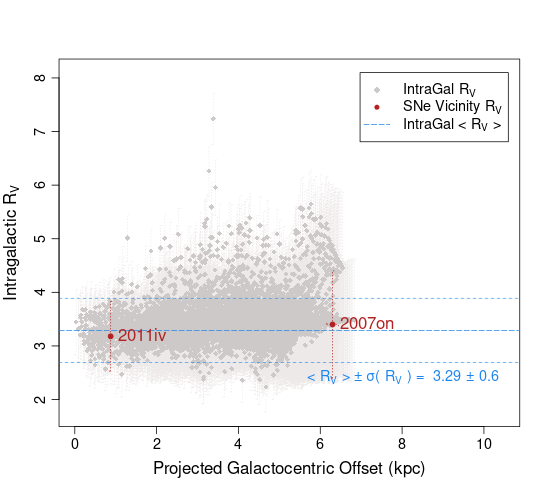}
    \caption{$R_V$ estimated for each $5\times5$ \SI{}{pix}$^2$ bin \textit{versus} projected galactocentric offset (grey diamonds), for NGC-1404. The blue dashed lines mark the global $R_V$, $<R_V>$, and the short dashed blue lines mark enclose the 1$\sigma$ uncertainty of the median. The red dot indicates our estimate/inference for the vicinity of the hosted SNe. The average uncertainty of $R_V$ across bins is 0.027 and each bin has a projected length of \SI{23}{pc}. For other galaxies in the sample, see~\ref{app:D}}
    \label{fig:IGM_RV_1404}
\end{figure}

Indeed, comparing the $R_V$ estimates at the SNe vicinities to the median\footnote[12]{Median of $5\times5$ bins well modeled by the Serkowski relation} $R_V$ of the HGs (provided in Tab. \ref{tab:HG_Rvs} and Fig. \ref{fig:Rv_SNxHG}), we observe that the estimates for all SNe vicinities fall below 1$\sigma$ of the HG median. Moreover, comparing to all other regions of a given galaxy (see e.g. Fig.~\ref{fig:IGM_RV_1404}), we see that the $R_V$ values in different locations span mostly values between $ \sim2.5-5$, as in the MW, but not much below that. This is true for all galaxies in our sample (see Appendix~\ref{app:D}). These results provide further evidence that not only is there nothing unusual at the extended site around the SNe, but that there are no regions in the galaxies that meet the peculiar low $R_V$ reddening laws found for some highly extincted SNe~Ia. This strengthens the hypothesis that either the average dust properties of more extended regions are not representative of more localized individual ISM clouds, or the SN explosion interacts with the nearby environment, altering its dust properties.

\begin{table*}
\centering
{\renewcommand{\arraystretch}{1.4}
\begin{tabular}{|c|c|c|c|c|}
\hline
Host Galaxy               & Median $R_V$                   & SN     & $R_V$         & $\Delta R_V(\sigma)$ \\ \hline
\multirow{2}{*}{NGC-1404} & \multirow{2}{*}{3.29$\pm$0.60} & 2007on & $3.4\pm1.0$ & 0.09           \\ \cline{3-5} 
                          &                                & 2011iv & $3.18\pm0.65$ & 0.13           \\ \hline
NGC-3244                  & 3.14$\pm$0.60                  & 2010ev & $3.16\pm0.65$ & 0.02           \\ \hline
NGC-3351                  & 2.92$\pm$0.55                  & 2012aw & $3.33\pm0.75$ & 0.44           \\ \hline
NGC-4424                  & 3.49$\pm$0.66                  & 2012cg & $3.50\pm0.82$ & 0.01           \\ \hline
\multirow{2}{*}{NGC-6962} & \multirow{2}{*}{3.01$\pm$0.55} & 2002ha & $3.04\pm1.28$ & 0.02           \\ \cline{3-5} 
                          &                                & 2003dt & -             & -              \\ \hline
NGC-7721                  & 3.05$\pm$0.51                  & 2007le & $3.28\pm0.51$ & 0.32           \\ \hline
NGC-7780                  & 3.04$\pm$0.56                  & 2001da & $3.98\pm0.36$ & 1.40           \\ \hline
UGC-11723                 & 3.08$\pm$0.59                  & 2006cm & $3.01\pm0.54$ & 0.09           \\ \hline
UGC-12859                 & 3.24$\pm$0.65                  & 2007fb & $3.15\pm0.58$ & 0.10           \\ \hline
\end{tabular}
}
\caption{Global $R_V$ for the present sample of SN HGs compared against the $R_V$ at the vicinity of the hosted SNe. The global $R_V$ is a median estimated using all bins well modeled by a Serkowski relation.}
\label{tab:HG_Rvs}
\end{table*}

\subsection{$R_V$ relations with galaxy properties}
\begin{figure}[ht]
    \centering
    \includegraphics[width =\linewidth]{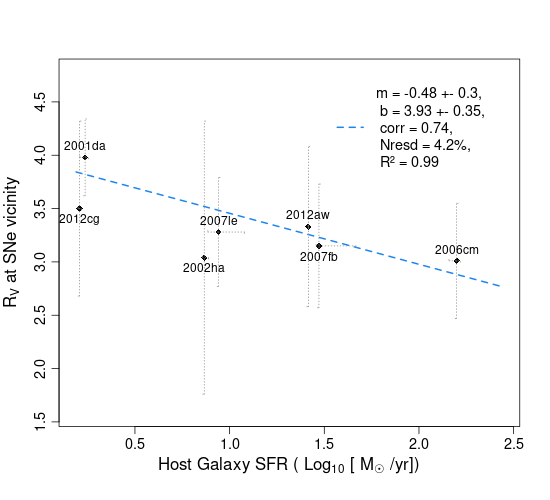}
    \caption{Our $R_V$ estimations at SNe vicinity \textit{versus} host galaxy global star formation ratio.}
    \label{fig:Rv_vs_props}
\end{figure}

Despite our small SN sample, we attempt to further our analysis into investigating the possible relationship between our measured $R_V$ and the properties of the SN host galaxies. We thus extracted and/or derived from the Asiago Supernova Catalogue\footnote{\url{http://graspa.oapd.inaf.it/asnc/}} and NED\footnote{\url{https://ned.ipac.caltech.edu/}}: morphology (Hubble T-type), normalized galactocentric distance and normalized directional light radius \citep{Gupta16}, distance to Earth, and galactic inclination. In addition, from stellar population fits to multi-band photometry carried out in \citet{santi2025}, we obtained the global, and local (\SI{0.5}{kpc} radius vicinity) stellar mass, stellar age, dust attenuation $A_V$, dust index, $R^{n}_V$\footnote{This $R_V$ was derived from the dust index and $A_V$ using the modified Calzetti attenuation law \citep{noll2009, kriek2013} and Eq. \ref{eq:RvRed}.}, star formation rate (SFR) and specific SFR, for 7 of the 9 galaxies in our data (all except for NGC-1404 and NGC-3244). Of all these properties, only the global SFR presents a correlation greater than 0.7 with our $R_V$ estimates at the SN location, as can be seen in Fig. \ref{fig:Rv_vs_props}. Despite large uncertainties in the parameters and a small sample, a decreasing trend is observed. Lower $R_V$ indicates the presence of smaller grains in regions with higher star formation rate, as has also been recently found in \cite{Soliman+24}, although we do not find it for the local SFR. 

Additionally, for each galaxy in our sample we select all bins whose fit to the Serkowski relation matched our filtering criteria (see Section \ref{sec:meth}), and compare the relationship between $R_V$ and the projected galactocentric offset. Again, we find no relationship between these parameters, despite the presence of some extreme outliers (illustrations of these results can be found in Appendix \ref{app:D}). Given that for photometric data fitted to stellar population models, the dust attenuation law is highly dependent on scattering and geometric effects and thus follows a decreasing trend with galactocentric distance (see Fig.~11-15 of \citealt{Duarte25} or Fig.~6 of \citealt{hutton15}), this is a reassuring result: polarimetric data carefully fitted to Serkowski laws is perhaps more appropriate to infer dust reddening than intensity data fitted to stellar population models. It provides a cleaner way to probe the real $R_V$ and the actual physical properties of the dust.

\section{Concluding remarks}
Using galaxy multi-waveband ($BVRI$) photopolarimetry we have analyzed 9 host galaxies of 11 SNe, most of which are highly extincted SNe~Ia with very peculiar dust reddening laws. By finding regions well fitted by a Serkowski polarization-wavelength relation and thus mainly driven by dust extinction and grain alignment (and not scattering), we obtain the maximum wavelength of polarization and with it the reddening law parameterized with $R_V$. By binning different regions and using spatial inference when necessary, we obtain $R_V$ in the SN vicinity and across the galaxies. We find no evidence of peculiar extinction laws in the 0.1-3 \SI{}{kpc} vicinity of these reddened SNe~Ia. The $R_V$ values are consistent with the average value of the MW and they are representative of the entire host galaxy; but are significantly different ($\sim3.5\sigma$) from the values obtained directly from the SN light around maximum. 

Our results match the $R_V$ values from host galaxy intensity analysis at the SN site before explosion \citep{hutton15}, as well as $R_V$ estimates from late-time SN observations (10-20 days after maximum light by \citealt{Amanullah2015} or after 35 days by \citealt{forster13}) for the same SNe~Ia. A viable explanation for this discrepancy is the interaction of the SN light with the immediate neighboring material, CSM or patchy ISM, which occurs for explosions in very dusty environments \citep{forster13,hoang17,bulla18}.

This study demonstrates the independent and powerful capabilities of photopolarimetry to obtain dust properties in various systems. Care should be taken with contamination from polarization due to scattering and/or the effects of multiple clouds with differing dust properties in the LOS \citep{mandarakas24}. The combination of spectro-polarimetry and multi-waveband photopolarimetry in the follow-up of SN events could provide crucial data to constrain the physical properties of the nearby material. Progress in the inclusion of polarization effects in radiative transfer codes is also necessary for more accurate modeling of these events.\par

\begin{acknowledgements}
João Rino-Silvestre was supported by FCT (PD/BD/150487/2019), through the IDPASC (\url{https://idpasc.lip.pt/}) PhD program. The work here described was supported by the CRISP project (PTDC/FIS-AST/31546/2017).\\
We would like to thank Takashi Nagao, for his insights on different features of the polarization of light by interactions with dust grains, and Maarten Baes for his help and instruction with \texttt{SKIRT}.\\
This research has used the NASA/IPAC Extragalactic Database (NED), which is funded by the National Aeronautics and Space Administration and operated by the California Institute of Technology.\\
This research has made use of the Asiago Supernova Catalogue \citep{asnc}.\\
This work has used data from the European Space Agency (ESA) mission {\it Gaia} (\url{https://www.cosmos.esa.int/gaia}), processed by the {\it Gaia} Data Processing and Analysis Consortium (DPAC, \url{https://www.cosmos.esa.int/web/gaia/dpac/consortium}). Funding for the DPAC has been provided by national institutions, in particular, the institutions participating in the {\it Gaia} Multilateral Agreement.
\end{acknowledgements}

\bibliography{bib}
\bibliographystyle{aa}

\appendix
\section{\texttt{SKIRT} models}\label{app:A}
Stellar Kinematics Including Radiative Transfer (\texttt{SKIRT})\citep{skirt1,skirt2} is a Monte Carlo radiative transfer suite that offers some built-in source templates, geometries, dust characterizations, spatial grids, and instruments, as well as an interface so that a user can easily describe a physical model. The user can in this way avoid coding the physics that describes both the source (e.g. AGN or galaxy type, observation perspective, emission spectrum) and environment (between the simulated source and observer, such as dust grain type and orientation, dust density distribution, etc.) but instead, design a model of modular complexity by following a Q\&A prompt (itself adaptable to the user expertise).\par

To demonstrate the spatial polarization angle pattern, displayed in Fig. \ref{fig:scatAng}, that signals scattering as the dominant process in polarizing the light that comes into the LOS, we constructed a simple, one source one medium model. The details of the model are given below, while a ready-to-run \texttt{SKIRT} input file and scripts to process the simulation outputs are \href{https://github.com/p41nk1ll3r/SKIRT_Scat_Pol/tree/main}{publicly available}.

The light source is modeled by a spherical Gaussian geometry of the form.

\begin{align}
    \rho (r) = \rho_0 \exp{\Bigg(-\frac{r^2}{2\sigma^2}\Bigg)}\:,
\end{align}

\noindent with $\sigma = $\SI{0.05}{kpc}, with a black body emission spectrum with temperature $T = $ \SI{5000}{K}. The medium is modeled by a homogeneous sphere with radius $r = $ \SI{0.5}{kpc} and by dust properties that replicate those of a mixture of silicates, graphites, and polycyclic aromatic hydrocarbons with composition and size distributions that accurately replicate the extinction, emission and abundances constraints for the Milky Way \citep{camps2015, gordon2017}, based on the \texttt{BARE‐GR‐S} dust model by \cite{zubko}.\par

In one simulation the total dust mass is normalized to achieve an optical depth\footnote{defined as the natural logarithm of the ratio between incident $F_I$, and transmitted $F_T$, radiation $\tau = \ln{\frac{F_I}{F_T}}$.}, at $\lambda = $\SI{0.35}{\um} and along length of \SI{1}{kpc}, of $\tau = 1$. In the second simulation, to increase the fraction of light reaching the LOS through multiple instances of scattering, the optical depth was increased to $\tau = 5$. 

\section{Choosing $R_V$ estimates}\label{app:Rv}

Following the fitting procedure to Serkowski laws detailed in section~\ref{sec:meth}, estimates were obtained for $R_V$ at the approximate locations of 5 of the 11 supernovae. This procedure failed to fit satisfying Serkwoski relations and thus recover reliable $R_V$ estimates for the vicinity of SNe hosted in NGC-3244, NGC-3351, NGC-4424, NGC-6962, and NGC-7721, as well as for the vicinity of SN 2007on in NGC-1404. In the cases of NGC-3351 and NGC-4424 this was expected considering that their polarization angle maps present (in some wavebands) strong indications of scattering as the dominant source of polarization (see Fig. \ref{fig:scatGals}). Surprisingly, however, looking at the percentage of bins successfully fitted and within all our criteria (percentage of success, or PoS) there appears to be no relation with that factor. The average PoS, across galaxies, is $25\pm5$\% while the PoS for NGC-3351 and NGC-4424 are 22\% and 25\%, respectively. However, the lowest PoS is for NGC-3244 at 18\% which does not display a scattering polarization signature in any of the observed wavebands.\par

As for the case of NGC-6962, the locations of SNe 2002ha and 2003dt unfortunately coincided with the gap between the two chips of the CCD. SN 2007on is located on the outskirts of NGC-1404, a region where the flux level, unfortunately, matched the background and was therefore rejected\footnote{Our reduction procedure corrects for background polarization by removing the estimated background flux, following \citep{patatrom}. More details are provided in the companion paper \citep{self-pipe}.}.\par

Although an estimate directly from the bin with the location of SN 2007le, in NGC-7721, was not possible, we were able to use two well fitted neighboring bins from which we took the mean of their $R_V$ values. The estimate $R_V$ for SN 2006cm is based on Serkowski model fit with only three bands (the \texttt{R\_SPECIAL} SNR for that bin is too low), while the number of parameters in the model is also three, making the result less reliable. The results are presented in Tab. \ref{tab:sne_Rv_vanila}. \par

\begin{table*}[ht]
\centering
{\renewcommand{\arraystretch}{1.25}
\begin{tabular}{|c|c|c|c|c|c|c|}
\hline
SN & Host Galaxy & $\lambda^{INLA}_{max}$ (\SI{}{\um}) & R$_V(\lambda^{INLA}_{max})$ & $R^{INLA}_V$ & $\Delta R_V(\sigma)$ & Projected Area (kpc$^2$)\\ \hline
2007on & NGC-1404 & $0.62\pm1.0$ & $3.4\pm5.5$ & $3.4\pm1.0$ & 0.00 & $0.12 \times 0.12$\\ \hline
2011iv & NGC-1404 & $0.62\pm0.48$ & $3.4\pm2.6$ & $3.18\pm0.65$ & 0.08 & $0.12 \times 0.12$\\ \hline
2010ev & NGC-3244 & $0.57\pm0.58$ & $3.1\pm3.2$ & $3.16\pm0.65$ & 0.02 & $0.17 \times 0.17$\\ \hline
2012aw & NGC-3351 & $0.61\pm0.67$ & $3.4\pm3.7$ & $3.33\pm0.75$ & 0.02 & $0.06 \times 0.06$\\ \hline
2012cg & NGC-4424 & $0.64\pm0.82$ & $3.5\pm4.5$ & $3.50\pm0.82$ & 0.00 & $0.09 \times 0.09$\\ \hline
2002ha & NGC-6962 & $0.55\pm1.28$ & $3.03\pm7.04$ & $3.04\pm1.28$ & 0.00 & $0.34 \times 0.34$\\ \hline
2007le & NGC-7721 & $0.59\pm0.36$ & $3.2\pm2.0$ & $3.28\pm0.51$ & 0.04 & $0.13 \times 0.13$\\ \hline
2001da & NGC-7780 & $0.72\pm0.23$ & $4.0\pm1.3$ & $3.92\pm0.38$ & 0.06 & $0.45 \times 0.45$\\ \hline
2006cm & UGC-11723 & $0.55\pm0.54$ & $3.0\pm3.0$ & $3.01\pm0.54$ & 0.00 & $0.43 \times 0.43$\\ \hline
2007fb & UGC-12859 & $0.57\pm0.59$ & $3.1\pm3.2$ & $3.15\pm0.58$ & 0.02 & $0.45 \times 0.45$\\ \hline
\end{tabular}
}
\caption{Estimates of $\lambda^{INLA}_{max}$, $R^{INLA}_V$ and R$_V(\lambda^{INLA}_{max})$ in the vicinity of the SNe hosted by the galaxies in our sample. These first two estimates were obtained by approximate spatial inference of the previously obtained bin maps of $\lambda_{max}$ and $R_V$ using \texttt{INLA}, while R$_V(\lambda^{INLA}_{max})$ was calculated using $\lambda^{INLA}_{max}$ map and Eq. \ref{eq:Rv}. The projected area, in \SI{}{kpc}$^2$, corresponding to the bin selection is also provided.}
\label{tab:sne_Rv_inla}
\end{table*}

In an attempt to address this and also improve some of our previous estimates on the remaining SNe locations, we employ \texttt{INLA} to perform spatial inference on the $\lambda_{max}$ and $R_V$ maps obtained in the previous step. As the uncertainty of the inference, we add in quadrature the \texttt{INLA} output standard deviation for the inference together with the median uncertainty of the input map. We perform equivalent inferences of the $R_V$ at SNe locations that had already been successfully estimated (as shown in Tab. \ref{tab:sne_Rv_vanila}). We find virtually no tension between the $R_V$ values obtained directly from spatial inference and those obtained by applying Eq. \ref{eq:Rv} to the spatial inferences performed on the $\lambda_{max}$ maps. However, the uncertainties for the $R_V$ values obtained from the inferred $\lambda_{max}$ maps are much larger\footnote{\texttt{INLA} standard deviation for inference is based on the value range of the data, and, the distances between the inferred point and the points used for the inference.} than the ones obtained by inference from the $R_V$ maps, for that reason we chose to use these last ones instead. For more details on the results obtained with \texttt{INLA} see Tab. \ref{tab:sne_Rv_inla}.\par

To verify the validity of the inferences made with \texttt{INLA}, the differences $\Delta R_V$, between the inferred $R_V$ values and those obtained directly with the Serkowski fits were calculated. The results, given in units of $\sigma$, are presented in Tab. \ref{tab:sne_Rv_obs_vs_inla}, show the $R_V$ to be within $1\sigma$ of the estimates based on bins with size $5\times5$ \SI{}{pix}$^2$ and with size $25\times25$ \SI{}{pix}$^2$. We then analyzed the SNe neighborhood in the $5\times5$ \SI{}{pix}$^2$ bin map, to decide whether to use the estimates based on the $25\times25$ \SI{}{pix}$^2$ bins or the inferences of $R_V$ obtained with \texttt{INLA}.\par

The bin mean $R_V$ estimated for the vicinity of SN 2007le is very similar to the value inferred with \texttt{INLA} and, despite a lack of $5\times5$ \SI{}{pix}$^2$ bins well fitted nearby, within the corresponding $25\times25$ \SI{}{pix}$^2$ bin area the estimates of $R_V$ are very similar. Therefore, we chose to take the inferred value, as it reports to a smaller area while taking into account information from its surroundings.\par

In the case of SN 2006cm the corresponding $5\times5$ \SI{}{pix}$^2$ bin estimate did not pass the selection criteria. In the neighborhood at a distance of 1 bin, we find only one reasonable estimate ($\lambda_{max} = 0.487\pm0.065$ \SI{}{\um}) and two outliers without credibility ($\lambda_{max} = {0.02\pm0.10, 0.1\pm4.5}$ \SI{}{\um}). For this reason, we prefer to use the $R_V$ estimated from the $25\times25$ \SI{}{pix}$^2$ bin over any estimate obtained from $5\times5$ \SI{}{pix}$^2$ bins. Looking further into the $5\times5$ \SI{}{pix}$^2$ bins that make up the $25\times25$ \SI{}{pix}$^2$ bin, we find 10 other successfully fitted bins, three of them without credibility ($\lambda_{max} = {1.63\pm0.73, 0.02\pm0.16, 0.01\pm0.24}$ \SI{}{\um}) and the remaining 7 with $R_V$ = {2.76, 3.13, 3.27, 3.37, 3.40, 3.52, 3.63} $\pm$ 0.36. The $R_V$ estimated from the $25\times25$ \SI{}{pix}$^2$ bin thus is not only the result of a less than desirable amount of polarimetry (no data available for that bin with the filter \texttt{R\_SPECIAL}) but also from a mixture of properties of the different spatial sections within. For these reasons, even though the $\Delta\sigma$ between the estimate and the \texttt{INLA} inference is below 1 and the inference presents a greater uncertainty, we chose to adopt the $R_V$ inferred by \texttt{INLA}.\par 

In the case of SN 2007fb the corresponding $5\times5$ \SI{}{pix}$^2$ bin did not pass or selection criteria either, and at a distance of 1 bin not a single credible estimate is found, only outliers with $\lambda_{max}$ well beyond the scope of application of the Serkowski relation which would physically imply a drastically different dust grain size distribution from what has been previously predicted by different models \citep{gald1, zubko} for dust in the ISM. The diversity of the neighborhood may be intrinsic in this case, since the center of each bin stands at $\sim$ \SI{200}{pc} from their closest neighbors (far beyond the $\ion{H}{II}$ region scales), or it may be a manifestation of poor SNR and/or resolution, where properties from different regions may be mixed. The estimates $R_V$ based on the $5\times5$ \SI{}{pix}$^2$ bins were then rejected due to a lack of credibility in the fit to the Serkowski relation in the SN bin and to the diversity of values in the neighborhood. However, the fact that the estimate for the $25\times25$ \SI{}{pix}$^2$ bin is incompatible with any of the bins that fit inside it points toward a mixture of properties due to a loss in spatial resolution, as mentioned above. For this reason, we chose to adopt the $R_V$ inferred by \texttt{INLA}. For the same reason, we also chose the $R_V$ inferred for SN 2011iv instead of the $25\times25$ \SI{}{pix}$^2$ bin estimate.\par

\begin{table}[ht]
\centering
\small
{\renewcommand{\arraystretch}{1.3}
\begin{tabular}{|c|c|c|c|c|}
\hline
SN & Host Galaxy & $R_V$ & $R^{INLA}_V$ & $\Delta R_V(\sigma)$\\ \hline
2011iv & NGC-1404 & $3.18\pm0.36$ & $3.18\pm0.65$ & 0.00 \\ \hline
2007le & NGC-7721 & $3.17\pm0.65$ & $3.28\pm0.51$ & 0.13 \\ \hline
2001da & NGC-7780 & $3.98\pm0.36$ & $3.92\pm0.38$ & 0.11 \\ \hline
2006cm & UGC-11723 & $2.62\pm0.36$ & $3.01\pm0.54$ & 0.60 \\ \hline
2007fb & UGC-12859 & $2.70\pm0.36$ & $3.15\pm0.58$ & 0.66 \\ \hline
\end{tabular}
}
\caption{Differences ($\Delta R_V$) between $R_V$ estimated from Serkowski fits and inferred with \texttt{INLA} from spatial maps of $R_V$ estimations, given in $\sigma$.}
\label{tab:sne_Rv_obs_vs_inla}
\end{table}

\section{Host Galaxy IGM $R_V$ \textit{versus} projected galactocentric offset}\label{app:D}
In the following figures we plot, for each galaxy in our sample, the total-to-selective extinction ratio \textit{versus} the projected galactocentric offset of each $5\times5$ \SI{}{pix}$^2$ bin which fit to a Serkowski relation matched all our filtering criteria (see Section \ref{sec:meth}). Unlike \citep{hutton15}, we do not find a relation between these two properties.

\begin{figure}[ht]
    \centering
    \includegraphics[width=\linewidth]{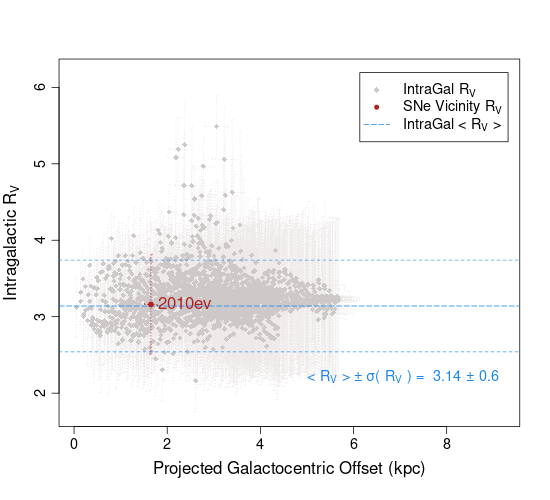}
    \caption{$R_V$ estimated for each $5\times5$ \SI{}{pix}$^2$ bin \textit{versus} projected galactocentric offset, for NGC-3244. Same as key as \ref{fig:IGM_RV_1404}. The average uncertainty of $R_V$ across bins is 0.057 and each bin has a projected length of \SI{34}{pc}.}
    \label{fig:IGM_RV_3244}
\end{figure}

\begin{figure}[ht]
    \centering
    \includegraphics[width=\linewidth]{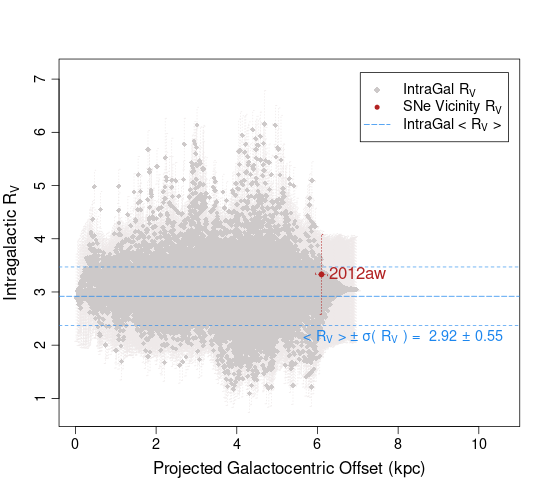}
    \caption{$R_V$ estimated for each $5\times5$ \SI{}{pix}$^2$ bin \textit{versus} projected galactocentric offset, for NGC-3351. Same as key as \ref{fig:IGM_RV_1404}. The average uncertainty of $R_V$ across bins is 0.029 and each bin has a projected length of \SI{12}{pc}.}
    \label{fig:IGM_RV_3351}
\end{figure}

\begin{figure}[ht]
    \centering
    \includegraphics[width=\linewidth]{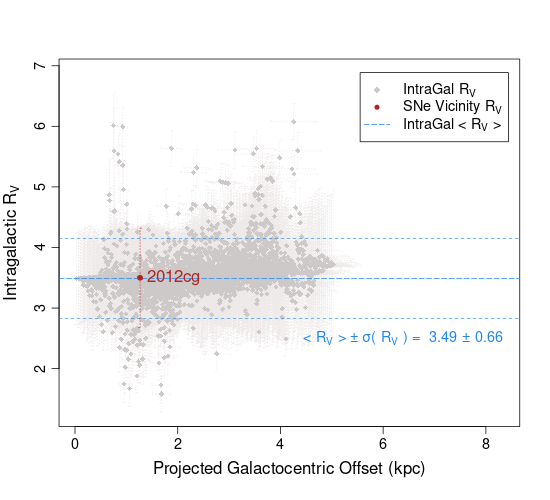}
    \caption{$R_V$ estimated for each $5\times5$ \SI{}{pix}$^2$ bin \textit{versus} projected galactocentric offset, for NGC-4424. Same as key as \ref{fig:IGM_RV_1404}. The average uncertainty of $R_V$ across bins is 0.034 and each bin has a projected length of \SI{17}{pc}.}
    \label{fig:IGM_RV_4244}
\end{figure}

\begin{figure}[ht]
    \centering
    \includegraphics[width=\linewidth]{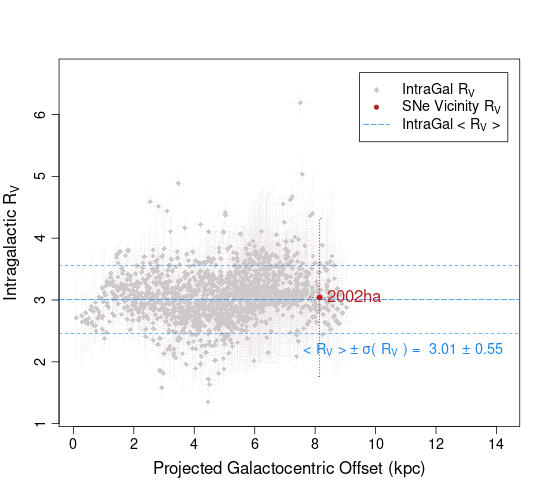}
    \caption{$R_V$ estimated for each $5\times5$ \SI{}{pix}$^2$ bin \textit{versus} projected galactocentric offset, for NGC-6962. Same as key as \ref{fig:IGM_RV_1404}. The average uncertainty of $R_V$ across bins is 0.023 and each bin has a projected length of \SI{68}{pc}.}
    \label{fig:IGM_RV_6962}
\end{figure}

\begin{figure}[ht]
    \centering
    \includegraphics[width=\linewidth]{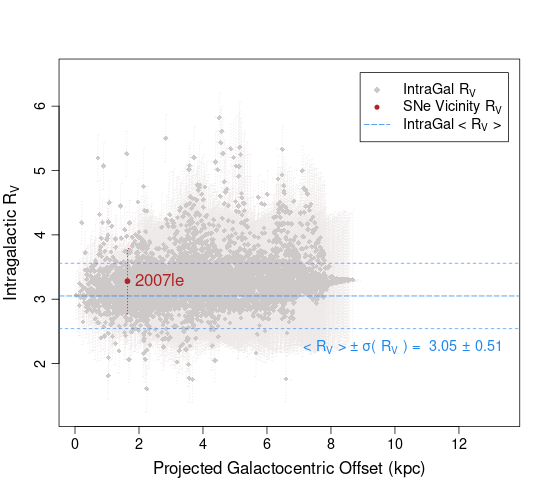}
    \caption{$R_V$ estimated for each $5\times5$ \SI{}{pix}$^2$ bin \textit{versus} projected galactocentric offset, for NGC-7721. Same as key as \ref{fig:IGM_RV_1404}. The average uncertainty of $R_V$ across bins is 0.023 and each bin has a projected length of \SI{25}{pc}.}
    \label{fig:IGM_RV_7721}
\end{figure}

\begin{figure}[ht]
    \centering
    \includegraphics[width=\linewidth]{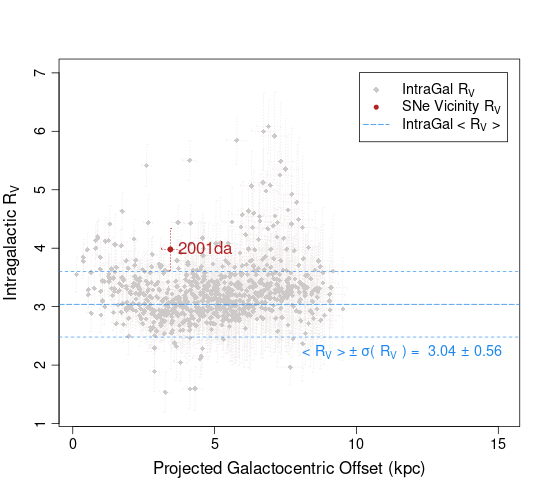}
    \caption{$R_V$ estimated for each $5\times5$ \SI{}{pix}$^2$ bin \textit{versus} projected galactocentric offset, for NGC-7780. Same as key as \ref{fig:IGM_RV_1404}. The average uncertainty of $R_V$ across bins is 0.022 and each bin has a projected length of \SI{90}{pc}.}
    \label{fig:IGM_RV_7780}
\end{figure}

\begin{figure}[ht]
    \centering
    \includegraphics[width=\linewidth]{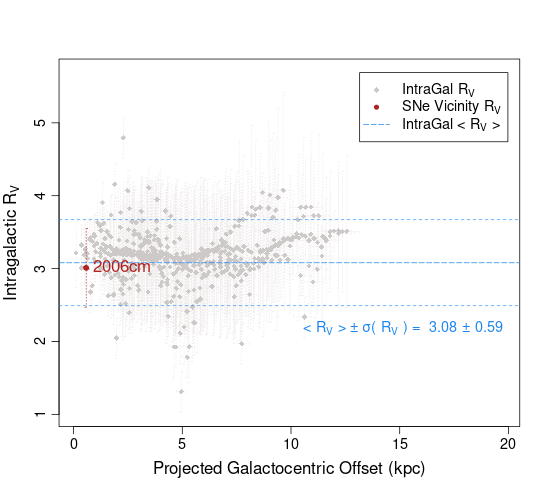}
    \caption{$R_V$ estimated for each $5\times5$ \SI{}{pix}$^2$ bin \textit{versus} projected galactocentric offset, for UGC-11723. Same as key as \ref{fig:IGM_RV_1404}. The average uncertainty of $R_V$ across bins is 0.017 and each bin has a projected length of \SI{87}{pc}.}
    \label{fig:IGM_RV_11723}
\end{figure}

\begin{figure}[ht]
    \centering
    \includegraphics[width=\linewidth]{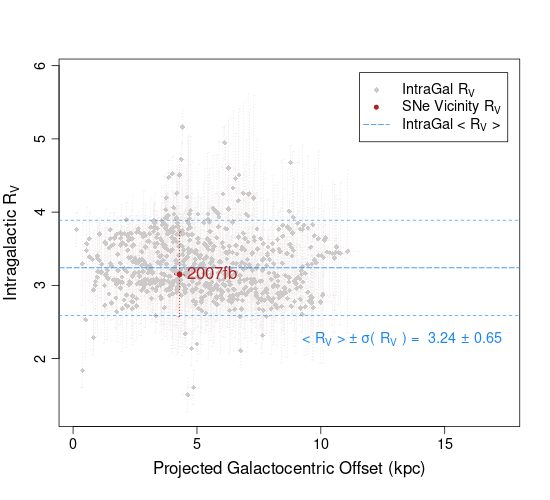}
    \caption{$R_V$ estimated for each $5\times5$ \SI{}{pix}$^2$ bin \textit{versus} projected galactocentric offset, for UGC-12859. Same as key as \ref{fig:IGM_RV_1404}. The average uncertainty of $R_V$ across bins is 0.021 and each bin has a projected length of \SI{91}{pc}.}
    \label{fig:IGM_RV_12859}
\end{figure}

\end{document}